\documentclass[aps,prb,twocolumn,showpacs,superscriptaddress,groupedaddress]{revtex4}

\usepackage{graphicx}
\usepackage{sidecap}
\usepackage{dcolumn}
\usepackage{bm}
\usepackage{amssymb}
\usepackage{amsmath}	
\usepackage{color}
\begin{document}


\widetext
\title{Tight-binding analysis of the electronic structure of\\ dilute bismide alloys of GaP and GaAs}


\author{Muhammad Usman} \email{usman@alumni.purdue.edu} \affiliation{Tyndall National Institute, Lee Maltings, Dyke Parade, Cork, Ireland}
\author{Christopher A. Broderick} \affiliation{Tyndall National Institute, Lee Maltings, Dyke Parade, Cork, Ireland}
\affiliation{Department of Physics, University College Cork, Cork, Ireland}
\author{Andrew Lindsay} \affiliation{Tyndall National Institute, Lee Maltings, Dyke Parade, Cork, Ireland}
\author{Eoin P. O'Reilly} \affiliation{Tyndall National Institute, Lee Maltings, Dyke Parade, Cork, Ireland}
\affiliation{Department of Physics, University College Cork, Cork, Ireland}
\vskip 0.25cm



\begin{abstract}
We develop an atomistic, nearest-neighbor $sp^{3}s^{*}$ tight-binding Hamiltonian to investigate the electronic structure of dilute bismide alloys of GaP and GaAs. Using this model we calculate that the incorporation of dilute concentrations of Bi in GaP introduces Bi-related defect states in the band gap, which interact with the host matrix valence band edge via a Bi composition dependent band anti-crossing (BAC) interaction. By extending this analysis to GaBi$_{x}$As$_{1-x}$ we demonstrate that the observed strong variation of the band gap $\left( E_{g} \right)$ and spin-orbit-splitting energy $\left( \Delta_{SO} \right)$ with Bi composition can be well explained in terms of a BAC interaction between the extended states of the GaAs valence band edge and highly localized Bi-related defect states lying in the valence band, with the change in $E_{g}$ also having a significant contribution from a conventional alloy reduction in the conduction band edge energy. Our calculated values of $E_{g}$ and $\Delta_{SO}$ are in good agreement with experiment throughout the investigated composition range ($x \leq 13$\%). In particular, our calculations reproduce the experimentally observed crossover to an $E_{g} < \Delta_{SO}$ regime at approximately 10.5\% Bi composition in bulk GaBi$_{x}$As$_{1-x}$. Recent x-ray spectroscopy measurements have indicated the presence of Bi pairs and clusters even for Bi compositions as low as 2\%. We include a systematic study of different Bi nearest-neighbor environments in the alloy to achieve a quantitative understanding of the effect of Bi pairing and clustering on the GaBi$_{x}$As$_{1-x}$ electronic structure.
\end{abstract}

\maketitle


\section{Introduction}


Highly mismatched semiconductor alloys such as GaN$_{x}$As$_{1-x}$ have attracted considerable interest both from a fundamental perspective and also because of their potential device applications \cite{Henini_1,Buyanova_1,Erol_1,Pettinari_1}. When a small fraction of As is replaced by N in GaAs, the band gap $\left( E_{g} \right)$ initially decreases rapidly, by $\approx$ 150~meV when 1\% of As is replaced by N \cite{Shan_1,Lindsay_1}. Similar behavior has been experimentally observed in GaBi$_{x}$As$_{1-x}$, where $E_{g}$ has been observed to initially decrease by $90$~meV \cite{Pacebutas_1,Alberi_1,Tixier_1, Yoshida_1} per \% of Bi replacing As in the alloy. In addition to this, recent experiments \cite{Batool_1,Fluegel_1} have also revealed the presence of a large bowing of the spin-orbit-splitting (SO) energy ($\Delta_{SO}$) with increasing Bi composition.

It has been recently shown\cite{Sweeney_1} that by increasing the Bi composition in GaBi$_{x}$As$_{1-x}$ to $\approx$ 11\% we enter an $E_{g} < \Delta_{SO}$ regime in the alloy. This regime is of interest for the design of highly efficient optoelectronic devices since it opens up the possibility of suppression of the non-radiative CHSH Auger recombination process, a loss mechanism which plagues the efficiency and dominates the threshold current in III-V multinary lasers operating in the telecommunication wavelength range\cite{Sweeney_1}.

The band gap reduction in GaN$_{x}$As$_{1-x}$ has been explained using a band anti-crossing (BAC) model \cite{Shan_1,Lindsay_1}. It is well established that replacing a single As atom by N introduces a resonant defect level above the conduction band edge (CBE) in GaAs. This occurs because N is considerably more electronegative and is also 40\% smaller than As. The interaction between the resonant N states and the CBE of the host GaAs matrix then accounts for the observed rapid reduction in $E_{g}$.

It has been proposed that replacing As by Bi could introduce a similar BAC effect \cite{Alberi_1}. Like N and As, Bi is also a group V element, which lies below Sb in the periodic table. Bi is  much larger than As, and is also less electronegative. It should therefore be expected that any Bi-related resonant defect levels should lie in the valence band (VB) and that, if an anti-crossing interaction occurs, it will occur between the Bi-related defect level and the valence band edge (VBE) of the GaAs matrix.

Recent x-ray absorption spectroscopy analysis of GaBi$_{x}$As$_{1-x}$ \cite{Ciatto_1} revealed that Bi atoms tend to form pairs and clusters in the alloy for $x \gtrsim$ 2\%. For this reason it is also of interest to explore the impact of Bi pairs and Bi clusters on the electronic structure of GaBi$_{x}$As$_{1-x}$. Such analysis requires atomistic modelling of the alloy since continuum methods such as the $\textbf{k} \cdot \textbf{p}$ and effective mass approximations are unable to predict the influence of such pairs and clusters on the electronic structure.

Although there is an increasing experimental interest in the structural \cite{Ciatto_1,Ferhat_1} and electronic\cite{Alberi_1,Tixier_1,Pacebutas_1,Alberi_2,Batool_1} properties of GaBi$_{x}$As$_{1-x}$, there have been comparatively few theoretical studies to date of this extreme semiconductor alloy. Previous theoretical investigations have employed the LDA+C \cite{Janotti_1, Deng_1}, $\textbf{k} \cdot \textbf{p}$\cite{Alberi_1}, and empirical pseudopotential\cite{Zhang_1} methods. These theoretical approaches have provided a range of insights, but it remains controversial as to whether the observed reduction in energy gap is due to an anti-crossing interaction \cite{Alberi_2} or is better explained using more conventional alloy and disorder models\cite{Deng_1}.

\begin{SCfigure*}
\centering
\includegraphics[width=0.65\textwidth]%
{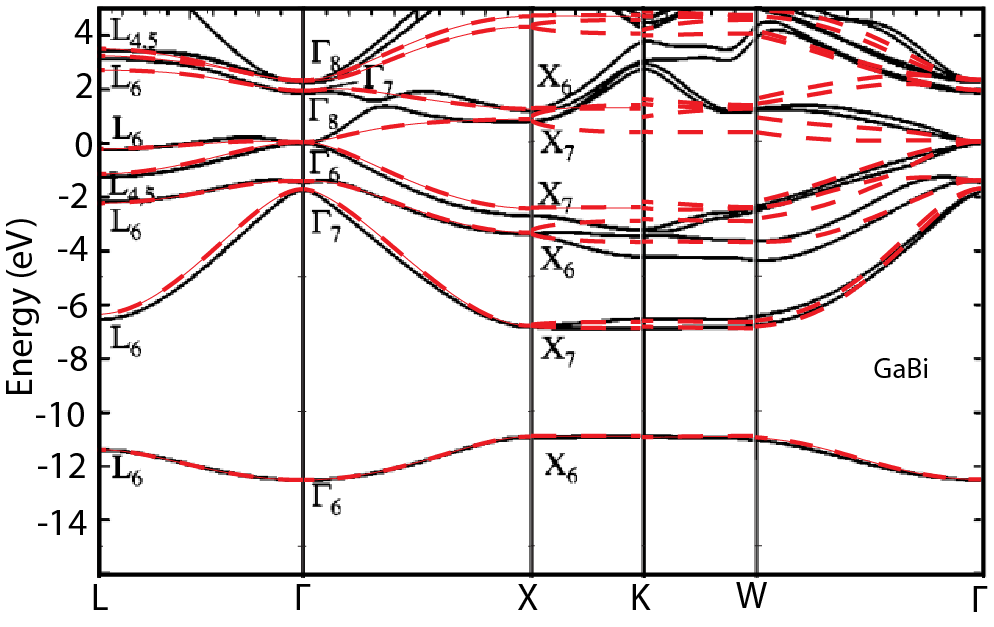}
\caption{ Bulk band structure of GaBi. The broken red lines are the bands obtained from our $sp^{3}s^{*}$ parametrisation. The solid black lines are the LDA+C bands calculated by the authors of Ref. 11. (used with permission)}
\label{fig:GaBi}
\end{SCfigure*}
\vspace{2mm}

The LDA+C calculations of Janotti  \emph{et al.} \cite{Janotti_1} presented the first band structure calculations of GaBi, and initial investigations of the influence of Bi on the electronic structure of GaBi$_x$As$_{1-x}$.  Zhang \emph{et al.} \cite{Zhang_1} presented pseudopotential calculations which were in good agreement with the experimentally measured values of $E_{g}$ and $\Delta_{SO}$ over the small composition range considered ($x \leq 4$\%). They also identified a Bi-related resonant defect state, $E_{Bi}$, which they found to lie 80~meV below the GaAs VBE in the dilute doping limit. The best choice of model to describe the  influence of such a resonant level on the band structure has recently  been discussed by Deng \emph{et al.}\cite{Deng_1}, who argued that it was not appropriate to use the BAC model to describe the impact of Bi incorporation, and proposed instead the use of a band-broadening picture. 

We find here a Bi-related level below the GaAs VBE in GaBi$_{x}$As$_{1-x}$, in agreement with previous work. Our comparative calculations of GaBi$_{x}$P$_{1-x}$ and GaBi$_{x}$As$_{1-x}$ confirm an anti-crossing interaction between the Bi-related resonant state and the VBE. We also  investigate the influence of Bi pair and cluster states on the electronic structure, showing that these pairs and clusters give rise to additional energy levels which tend to move into the energy gap with increasing cluster size, analogous to the behavior of N pair and cluster levels in GaAs\cite{Kent_1} and GaP\cite{Harris_1}.

The $\textbf{k} \cdot \textbf{p}$ calculations of Alberi \emph{et al.}\cite{Alberi_1} include explicitly BAC interactions between the GaAs VBE and lower-lying Bi-related levels in order to fit the experimentally observed variation of $E_{g}$ for dilute Bi compositions. They also propose a second BAC interaction between the GaAs SO band and a Bi-related SO defect level, 1.9 eV below the GaAs VBE. The inclusion of such a level would necessarily produce a BAC-induced upward shift in the SO band edge energy, $E_{SO}$. We find in our calculations that the evolution of the spin-split-off band is well described using a conventional alloy model, without the need to include such an SO-related BAC interaction.

We have previously demonstrated that a parametrised tight-binding (TB) approach provides an accurate and versatile method to analyse the electronic structure of highly mismatched semiconductor alloys such as GaN$_{x}$As$_{1-x}$, GaBi$_x$P$_{1-x}$, and GaBi$_{x}$As$_{1-x}$. Since a basis of localized states is employed,  it is possible using the TB method to probe explicitly and quantitatively the effect of an isolated impurity atom (e.g. replacing a single As atom by either N or Bi in GaAs) on the host matrix electronic structure\cite{Lindsay_2, Reilly_1}. The $sp^{3}s^{*}$ TB Hamiltonian developed for dilute nitride alloys\cite{Reilly_3} fully accounts for a wide range of experimental data, and also gives results in good agreement with pseudopotential calculations \cite{Kent_1}. We present here an $sp^{3}s^{*}$ TB Hamiltonian to calculate the electronic structure of dilute bismide alloys of GaP and GaAs. This Hamiltonian gives good agreement with pseudopotential and LDA+C calculations \cite{Zhang_1}, and with experiment over the full composition range investigated ($x \leq 13$\%).

We start our calculations with the study of ordered GaBi$_{x}$As$_{1-x}$ and GaBi$_{x}$P$_{1-x}$ supercells. It is useful to include GaBi$_{x}$P$_{1-x}$ in our analysis because Bi is sufficiently different to P that it introduces defect states just above the VBE in GaP \cite{Trumbore_1}. Since these Bi-related defect states are in the energy gap, it is then much easier to identify and follow the interaction between the VBE and these levels as a function of increasing Bi composition. Our calculations show clearly for GaBi$_{x}$P$_{1-x}$ that there is an anti-crossing interaction between the Bi defect states and the GaP VBE \cite{Reilly_2}. 

Having established that there is a BAC interaction in GaBi$_{x}$P$_{1-x}$, we then extend our calculations to consider ordered supercells of GaBi$_{x}$As$_{1-x}$. By comparison with GaBi$_{x}$P$_{1-x}$, we identify that there are Bi-related resonant defect levels below the VBE in GaBi$_{x}$As$_{1-x}$ and that these resonant defect levels interact with and push the VBE up in energy through an anti-crossing interaction. The effect of the resonant levels has been confirmed in GaN$_{x}$As$_{1-x}$ through photoreflectance (PR) measurements, which show a higher energy feature, $E_{+}$, associated with transitions to the resonant level\cite{Shan_1,Perkins_1,Klar_1}. We show that, because of the large density of valence states, it is not possible to experimentally identify similar higher energy transitions at any Bi composition in GaBi$_{x}$As$_{1-x}$.

Next, we investigate the effect of Bi pairs and clusters in GaBi$_{x}$As$_{1-x}$ alloys, by placing a single pair or cluster in a large (4096 atom) supercell in order to investigate its effect on the electronic structure. Our calculations highlight the evolution of the defect energy levels and their interaction with the VBE with increasing cluster size. This analysis later aids our understanding of the calculated electronic properties of randomly disordered GaBi$_{x}$As$_{1-x}$ supercells, in which the number of Bi pairs increases as  $x^2$ and the number of $n$-atom clusters increases as $x^n$.

Having established an understanding of the evolution of the Bi resonant state in ordered supercells containing only a single Bi atom, single Bi pair, or small Bi cluster, we next present calculations of disordered 4096 atom GaBi$_x$As$_{1-x}$ supercells with larger Bi compositions ($x \leq 13$\%). In such systems, we find that the CBE energy decreases both linearly and rapidly with increasing Bi composition, so that the experimentally observed reduction in the band gap is then due both to a BAC interaction in the VB and to conventional alloy behavior in the conduction band (CB). In addition, our calculations indicate that the observed strong bowing of the SO-splitting energy \cite{Fluegel_1, Batool_1} is predominantly attributable to the upward shift in the GaBi$_{x}$As$_{1-x}$ VBE with Bi composition, with a minor contribution from a downward shift in $E_{SO}$, which is linear in Bi composition and hence well understood as a conventional alloying effect. This linear variation in $E_{SO}$ is contrary to the SO-BAC interaction included in the $\textbf{k} \cdot \textbf{p}$ Hamiltonian introduced by Alberi \emph{et al.}\cite{Alberi_1} and shows that the Bi-related resonant state introduced in the VB interacts solely with the GaAs VBE at the Brillouin zone center ($\Gamma$ point).

Our calculations accurately reproduce the experimentally observed\cite{Batool_1} crossing of the band gap and the SO-splitting energies at a Bi composition of $\approx 10.5$\% in the alloy. Our analysis therefore lends support to the experimental conclusion that GaBi$_{x}$As$_{1-x}$ alloys can be used to obtain emission in the technologically important 1.3$\mu$m and 1.5$\mu$m wavelength ranges on a GaAs substrate, while also demonstrating that the unusual material behavior present in the alloy can be understood simply in terms of a BAC interaction.

The remainder of this paper is organised as follows: In section II we give a detailed presentation of the TB Hamiltonian and calculation procedure. In section III-A we apply the TB model to explore in detail the electronic structure of dilute ordered bismide alloys of GaP and GaAs, beginning with GaBi$_{x}$P$_{1-x}$. Following this, we continue our analysis of GaBi$_{x}$As$_{1-x}$ by performing a systematic study of the effects of Bi pairing and clustering on the electronic structure in section III-B. We then consider disordered GaBi$_{x}$P$_{1-x}$ supercells in section III-C and disordered GaBi$_{x}$As$_{1-x}$ supercells in section III-D. Finally we present our conclusions in section IV.


\section{Methodology}

\subsection{ $sp^{3}s^{*}$ tight-binding model}

We use an $sp^{3}s^{*}$ nearest-neighbor tight-binding Hamiltonian including spin-orbit coupling  to investigate the electronic structure of the different alloys. The inter-atomic interaction parameters vary with bond length and the magnitude of the on-site parameters depends explicitly on the overall neighbor environment \cite{Reilly_3}.

The alloy band structure calculations are undertaken using the \underline{N}ano\underline{E}lectronic \underline{MO}delling (NEMO 3-D) simulator \cite{Klimeck_1,Klimeck_2,Klimeck_3}. NEMO 3-D is a fully atomistic simulator and its capabilities include modelling of realistic multi-million atom geometries for embedded quantum dot devices \cite{Usman_1,Usman_2,Usman_3}, strained disordered quantum wells \cite{Karache_1}, and single impurities in Si FinFET structures\cite{Lansbergen_1}.

In our $sp^{3}s^{*}$ model, the on-site orbital energies for a given atom are determined by taking the average of the values from the binary compounds formed by that atom and its four nearest neighbors, taking into account the relevant valence band offsets (VBOs). The inter-atomic interaction energies are taken to vary with bond length, $d$, as $\left( \frac{d_{0}}{d} \right)^{\eta}$, where $d_{0}$ is the equilibrium bond length in the equivalent binary compound, and $\eta$ is a scaling parameter whose magnitude depends on the type of interaction being considered. The scaling indices for GaP and GaAs are determined by fitting the bond length dependence of their nearest-neighbor interaction parameters to a range of hydrostatic deformation potentials \cite{Reilly_1}. Since there are no experimental or theoretical deformation potentials available for GaBi, the scaling indices for GaBi were adapted from those of GaAs. We modify the GaAs $\eta$ values, in particular for the p states, in order to ensure that Bi gives a defect state in the energy gap of GaP, as observed experimentally\cite{Trumbore_1}. These modified values of $\eta$ for GaBi then give a Bi resonant state inside the valence band of GaAs for GaBiAs, consistent with earlier pseudopotential calculations\cite{Zhang_1}. The effects of changes in bond angle are taken into account through the two-center integral expressions of Slater and Koster \cite{Koster_1}.

\begin{figure}
\includegraphics[scale=0.55]{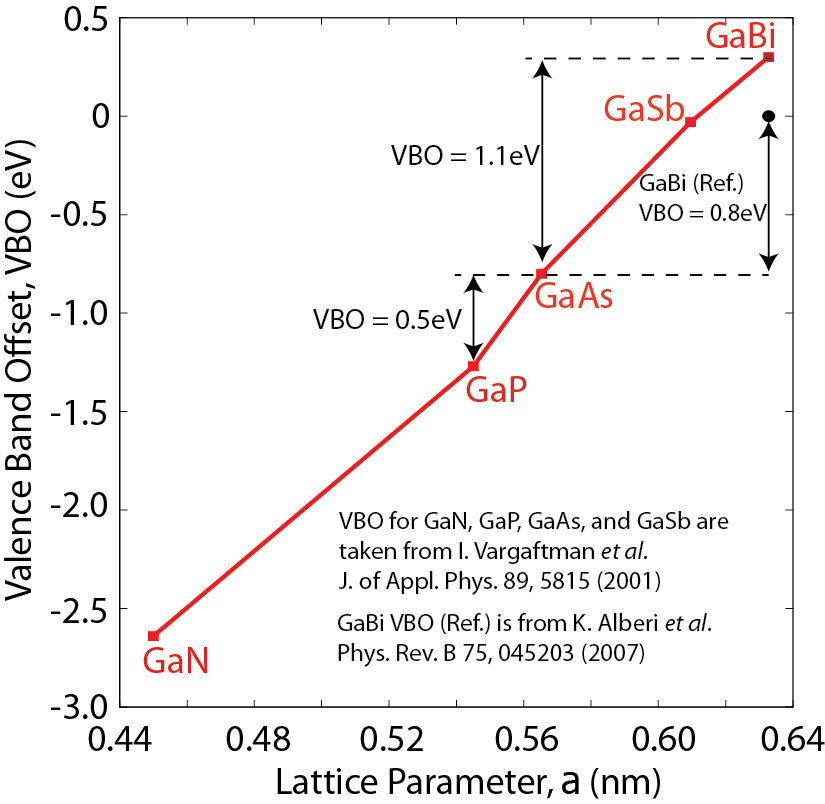}
\caption{Valence band offset (VBO) as a function of lattice parameter for Ga-containing III-V semiconductor binaries. The values for GaN, GaP, GaAs and GaSb are taken from Ref. {38}. The GaBi VBO is estimated by extrapolating the trend to the GaBi lattice constant (6.328 \AA) \cite{Francoeur_1}. For comparison, the GaBi VBO value of 0.8 eV estimated by the authors of Ref. 7 is indicated by a black dot.}
\label{fig:VBO}
\end{figure}
\vspace{1mm}

The self-energies and inter-atomic interaction parameters for the different binary compounds are listed in Table ~\ref{tab:table1}. The parameters for GaAs and GaP are based on those which we have used previously when studying GaN$_{x}$As$_{1-x}$ \cite{Reilly_3} and GaN$_{x}$P$_{1-x}$ \cite{Harris_1}, but we now include spin-orbit interactions. The parameters for GaBi were obtained by fitting to LDA+C calculations \cite{Janotti_1}. Fig.~\ref{fig:GaBi} shows the bulk GaBi band structure obtained from our $sp^{3}s^{*}$ Hamiltonian, compared with the LDA+C calculations of Janotti \emph{et al} \cite{Janotti_1}.

The valence band offsets for GaAs and GaP are taken from Ref. {38}. Since the binary compound GaBi has yet to be isolated, we need to estimate a value for the GaBi VBO. A VBO of 1.1eV for GaBi was estimated by an extrapolation of the VBO as a function of lattice parameter ($a$) for Ga-containing III-V semiconductor binaries (see Fig.~\ref{fig:VBO}). This differs from the value of 0.8eV used earlier by Alberi \emph{et al.}\cite{Alberi_1}, which appears to have been estimated by a linear extrapolation of VBO against the lattice parameter for GaN and GaP only. The scaling exponents, $\eta$, used to describe the variation of the inter-atomic matrix elements with bond length are listed in Table ~\ref{tab:table2}.

\begin{table}
\caption{\label{tab:table1}Lattice constants, orbital self-energies, inter-atomic interaction parameters, spin-orbit couplings and valence band offsets for GaP, GaAs and GaBi in the $sp^{3}s^{*}$ Hamiltonian. Lattice constants are given here in \AA; all other values are in eV.}
\begin{ruledtabular}
\begin{tabular}{cccc}
 & GaP & GaAs & GaBi \\
\hline
$a$ & 5.4505 & 5.6532 & 6.328 \\
$E_{s,a}$ & -8.1124 & -8.6336 & -8.3774 \\
$E_{s,c}$ & -2.1976 & -2.9474 & -5.6126 \\
$E_{p,a}$ & 1.0952 & 0.9252 & -0.1256 \\
$E_{p,c}$ & 4.0851 & 3.5532 & 1.694 \\
$E_{s^{*},a}$ & 8.4796 & 7.0914 & 6.1262 \\
$E_{s^{*},c}$ & 7.1563 & 6.2 & 5.8164 \\
$ss \sigma$ & -1.8727 & -1.6835 & -1.3425 \\
$pp \sigma$ & 3.0989 & 2.9500 & 2.0003 \\
$pp \pi$ & -0.7424 & -0.7420 & -0.6354 \\
$s_{a}p_{c} \sigma$ & 1.8500 & 2.3920 & 2.3567 \\
$s_{c}p_{a} \sigma$ & 2.7312 & 2.4200 & 1.2025 \\
$s_{a}^{*}p_{c} \sigma$ & 1.9998 & 2.0400 & 1.8051 \\
$s_{c}^{*}p_{a} \sigma$ & 2.1882 & 1.7700 & 0.5100 \\
$\lambda_{a}$ & 0.0666 & 0.405 & 2.0142 \\
$\lambda_{c}$ & 0.1734 & 0.165 & 0.1152 \\
VBO & -0.5 & 0 & 1.1
\end{tabular}
\end{ruledtabular}
\end{table}

\begin{table}
\caption{\label{tab:table2} Dimensionless scaling parameters, $\eta$, for the $sp^{3}s^{*}$ inter-atomic interaction parameters in GaP, GaAs and GaBi.}
\begin{ruledtabular}
\begin{tabular}{cccc}
 & GaP & GaAs & GaBi \\
\hline
$\eta_{ss \sigma}$ & 3.512 & 3.512 & 3.66 \\
$\eta_{sp \sigma}$ & 4.3 & 4.3 & 4.08 \\
$\eta_{pp \sigma}$ & 3.2042 & 3.2042 & 2.2 \\
$\eta_{pp \pi}$ & 4.2357 & 4.2357 & 3.24 \\
$\eta_{s^{*}p \sigma}$ & 5.5 & 5.5 & 5.5
\end{tabular}
\end{ruledtabular}
\end{table}

\begin{table}
\caption{\label{tab:table3}VFF parameters and coefficients for anharmonic corrections to the Keating potential for unstrained, bulk GaP, GaAs and GaBi. The VFF parameters $\alpha_{0}$ and $\beta_{0}$ are in units of N m$^{-1}$. The anharmonicity constants $A$, $B$ and $C$ are dimensionless.}
\begin{ruledtabular}
\begin{tabular}{cccc}
 & GaP & GaAs & GaBi \\
\hline
$\alpha_{0}$ & 44.5 & 41.49 & 34.0 \\
$\beta_{0}$ & 10.69 & 8.94 & 5.0 \\
$A$ & 7.2 & 7.2 & 8.2 \\
$B$ & 7.62 & 7.62 & 7.62 \\
$C$ & 6.4 & 6.4 & 6.4
\end{tabular}
\end{ruledtabular}
\end{table}

The TB calculations were carried out on supercells containing up to 4096 atoms using the NEMO 3-D simulator. For the disordered structures, Ga$_{2048}$Bi$_{M}$Y$_{2048-M}$ (Y = P, As) supercells were constructed by substituting Bi atoms at random Y sites on a GaY lattice and relaxing the crystal to its lowest energy configuration using a parametrised valence force field (VFF) model including anharmonic corrections \cite{Lazarenkova_1} to the Keating potential \cite{Keating_1}. The VFF parameters used are listed in Table ~\ref{tab:table3}. The parameters for GaAs are those obtained by Lazarenkova \cite{Lazarenkova_1}, while those for GaP and GaBi were derived based on the expressions of Martin \cite{Martin_1} and the elastic constants of Vurgaftman \emph{et al.} \cite{Vurgaftman_1} and Ferhat and Zaoui \cite{Ferhat_1}, respectively. It is worth noting that in the GaBi$_{x}$As$_{1-x}$ case relaxed Ga-Bi bond lengths from our model follow the same trend revealed by x-ray absorption spectroscopy measurements\cite{Ciatto_1}, where we see a decreasing Ga-Bi bond length in the presence of Bi pairing and clustering, indicating increased localized strain in the alloy.


\subsection{Tight-binding implementation of the band anti-crossing model}

The BAC model explains the extreme band gap bowing observed in Ga(In)N$_{x}$As$_{1-x}$ in terms of an interaction between two levels, one at energy $E_{M}$ associated with the extended CBE state $\vert \psi_{c,0} \rangle$ of the Ga(In)As matrix, and the other at energy $E_{D}$ associated with the localized N impurity defect states $\vert \psi_{N} \rangle$, with the two states linked by a N-composition dependent matrix element $V_{MD}$ describing the interaction between them \cite{Shan_1}. The CBE energy of Ga(In)N$_{x}$As$_{1-x}$ is then given by the lower eigenvalue, $E_{-}$, of the 2-band Hamiltonian 

\begin{equation}
      \left( \begin{array}{cc}
      E_{D} & V_{DM} \\
      V_{DM} & E_{M} \end{array} \right)
\end{equation}

A resonant feature associated with the upper eigenvalue, $E_{+}$, of Eq.~(1) has been observed in PR measurements \cite{Shan_1,Perkins_1,Klar_1}, appearing in GaN$_{x}$As$_{1-x}$ for $x \gtrsim 0.2$\% and remaining a relatively sharp feature until x $\approx$ 3\%, beyond which composition it broadens and weakens, when the resonant state becomes degenerate with the large $L$-related conduction band density of states \cite{Lindsay_3}.

To investigate the resonant state $\vert \psi_{N} \rangle$, and its behavior, we used the $sp^{3}s^{*}$ TB Hamiltonian to calculate the electronic structure of ordered GaN$_{x}$As$_{1-x}$ supercells \cite{Reilly_3,Lindsay_4}. By comparing the calculated host and alloy CBE states, $\vert \psi_{c,0} \rangle$ and $\vert \psi_{c,1} \rangle$, in large supercells (Ga$_{864}$N$_{1}$As$_{863}$ and Ga$_{864}$As$_{864}$, respectively), we derived the nitrogen resonant state, $\vert \psi_{N} \rangle$, associated with an isolated N atom. In the BAC model, $\vert \psi_{c,1} \rangle$ is a linear combination of $\vert \psi_{c,0} \rangle$ and $\vert \psi_{N} \rangle$, with $\vert \psi_{N} \rangle$ then given by:

\begin{equation}
      \vert \psi_{N} \rangle = \dfrac{\vert \psi_{c,1} \rangle -  \alpha \vert \psi_{c,0} \rangle}{ \sqrt{ 1 - \vert \alpha \vert^{2} }}
\end{equation}

\noindent
where, $\alpha = \langle \psi_{c,0} \vert \psi_{c,1} \rangle$. The evolution of the conduction band structure in an ordered GaN$_{x}$As$_{1-x}$ supercell can then be very well described by associating a localized resonant state with each N atom in the supercell\cite{Lindsay_2}.

In the dilute bismide case the BAC interaction is expected to occur in the valence band. Previous analysis\cite{Alberi_1} has included a BAC interaction with each of the three doubly degenerate valence bands, namely the heavy-hole (HH), light-hole (LH) and split-off (SO) bands. We find below that there is a BAC interaction with the HH and LH bands, but that there is no calculated BAC interaction involving the SO band. For dilute bismides, the Hamiltonian analogous to (1) is therefore:

\begin{equation} \left( \begin{array}{cccc}
 E_{HH} & 0 & V_{Bi} & 0 \\
 0 & E_{LH} &  0 & V_{Bi} \\
 V_{Bi} & 0 & E_{Bi} & 0 \\
 0 & V_{Bi} & 0 & E_{Bi} \\ \end{array} \right)
\end{equation}

\noindent
where $E_{HH/LH}$ are the (degenerate) host matrix VBE energies, $E_{Bi}$ is the energy of the Bi-related resonant defect state and $V_{Bi} = \beta \sqrt{x}$ is the composition-dependent interaction matrix element linking $E_{HH/LH}$ and $E_{Bi}$. For $E_{Bi}$ above (below) the host matrix VBE the interaction pushes down (up) the VBE.

Because of the 4-fold degeneracy of the VBE states, we need to modify the approach used in Eq.~(2)  to determine the BAC interaction in GaBi$_{x}$Y$_{1-x}$ (Y = P, As). As a result of the HH/LH degeneracy, the calculated alloy VBE states each include an arbitrary admixture of the host matrix VBE  states, $\vert \psi_{l,0} \rangle$. We can therefore use the calculated alloy and host matrix TB wave functions to determine the four Bi resonant states  $\vert \psi_{Bi,i} \rangle$ as a linear combination of the GaBi$_{x}$Y$_{1-x}$ VBE, $\vert \psi_{v,1,i} \rangle$, and $\vert \psi_{v,0,i} \rangle$, the part of the host (GaY) matrix VBE with which the $i$th Bi state, $\vert \psi_{Bi,i} \rangle$ interacts:

\begin{equation}
      \vert \psi_{Bi,i} \rangle = \dfrac{\vert \psi_{v,1,i} \rangle - \sum \limits^{4}_{l=1} \vert \psi_{l,0} \rangle \langle \psi_{l,0} \vert \psi_{v,1,i} \rangle}{ \sqrt{ 1 - \sum \limits^{4}_{l=1} \vert \langle \psi_{l,0} \vert \psi_{v,1,i} \rangle \vert^{2} }}
\end{equation}

\noindent
in which the index $l$ runs over the four-fold degenerate levels of the GaY VBE.

In general, for an isolated Bi impurity, $\vert \psi_{Bi,i} \rangle$ exhibits T$_d$ symmetry so it will be four-fold degenerate, with energy E$_{Bi}$. We can calculate explicitly $E_{Bi}$ and the interaction between $\vert \psi_{Bi,i} \rangle$ and the GaY VBE, $\beta$, by calculating the appropriate matrix elements of the full TB Hamiltonian for the Bi-containing alloy, $\widehat{H}$, for any one of the four degenerate Bi states:

\begin{eqnarray}
      E_{Bi} &=& \langle \psi_{Bi} \vert \widehat{H} \vert \psi_{Bi} \rangle \\
      \beta &=& \frac{V_{Bi}}{\sqrt{x}} = \frac{\langle \psi_{Bi} \vert \widehat{H} \vert \psi_{v,0} \rangle}{\sqrt{x}}
\end{eqnarray}

\noindent
where $\vert \psi_{v,0,i} \rangle = \sum \limits^{4}_{l=1} \vert \psi_{l,0} \rangle \langle \psi_{l,0} \vert \psi_{v,1,i} \rangle$, with the sum again running over the four-fold degenerate levels of the GaY VBE.


\subsection{Calculation of fractional $\Gamma$ character spectra in alloy supercells}

In the ideal BAC model of Eqs. (1) and (3), the host matrix band edge states are expected to mix in the alloy with the localized Bi-related defect states. In order to explore this BAC interaction, we calculate $G_{\Gamma} (E)$, which is the projection of the host matrix band edge states onto the full spectrum of levels in the alloy supercell. In what follows, we use the subscripts $l,0$ and $k,1$ to index the unperturbed host and Bi-containing alloy states, respectively. The $G_{\Gamma} (E)$ spectrum is calculated by projecting the unperturbed Ga$_{N}$Y$_{N}$ (Y = P, As) $\Gamma$ states, $\vert \psi_{l,0} \rangle$, onto the spectrum for the Ga$_{N}$Bi$_{M}$Y$_{N-M}$ alloy supercell, $\left\lbrace E_{k}, \vert \psi_{k,1} \rangle \right\rbrace$:

\begin{align}
	G_{\Gamma} \left( E \right) &= \sum_{k} \sum_{l=1}^{g(E_{l})} f_{\Gamma, kl} \; T \left( E_{k} - E \right) \\	
	f_{\Gamma, kl} &= \vert \langle \psi_{k,1} \vert \psi_{l,0} \rangle \vert ^{2}
\end{align}

\noindent
Here we choose $T \left( E_{k} - E  \right)$ as a \textquoteleft top hat\textquoteright \; function of width 2~meV and unit height centered on $E_{k}$, and $g(E_{l})$ is the degeneracy of the host (GaY) band having energy $E_{l}$ at $\Gamma$ (i.e. $g(E_{l}) = 2, 4$ and 2 for the conduction, valence and SO bands, respectively).

We use $G_{\Gamma} (E)$ firstly to confirm the presence of a BAC interaction in ordered GaBi$_{x}$P$_{1-x}$ and GaBi$_{x}$As$_{1-x}$ structures, and then to analyse the effects of Bi pair and cluster states, as well as the evolution of the BAC interaction in disordered structures. We also use $G_{\Gamma} (E)$ to explore the evolution of the conduction and SO band edges in GaBi$_{x}$As$_{1-x}$ with Bi composition, showing that the variation in both cases is attributable to the conventional alloying effect. These calculations are discussed in detail in the next section.

\section{Results and discussions}


\subsection{Bi state in ordered GaBi$_{x}$P$_{1-x}$ and GaBi$_{x}$As$_{1-x}$ supercells}

\subsubsection{Ga$_N$Bi$_1$P$_{N-1}$ supercells}

We have found previously that it is straightforward to identify and analyse a band anti-crossing interaction using Eqs. (4) to (7) in ordered supercell structures where the localized defect state is in the energy gap, as was the case for GaN$_{x}$P$_{1-x}$\cite{Harris_1}.  It has been shown experimentally that an isolated Bi atom acts as an impurity in GaP, giving rise to a defect level $\sim 100$ meV above the VBE \cite{Trumbore_1}. Our tight binding model has been established to give a Bi defect state in the band gap of GaP, consistent with this experimental study. We therefore start our calculations by investigating the electronic structure of ordered Ga$_N$Bi$_1$P$_{N-1}$ supercells, in which a single substitutional Bi atom is inserted into the GaP matrix. Having established the main factors influencing the valence band structure in GaBi$_{x}$P$_{1-x}$, we then consider GaBi$_{x}$As$_{1-x}$, using the phosphide results to guide and support the interpretation of the arsenide calculations.

\begin{figure}
    \includegraphics[scale=0.63]{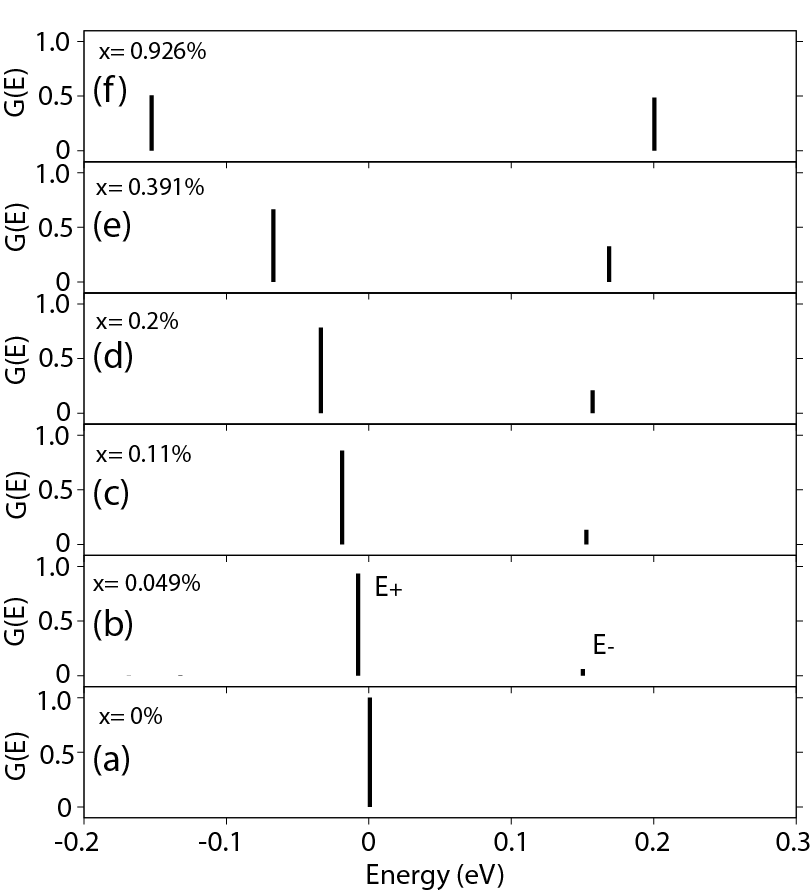}
 \caption{Combined fractional HH/LH $\Gamma$ character for a series of ordered Ga$_{N}$Bi$_{1}$P$_{N-1}$ supercells. The zero of energy is taken at the GaP VBE.}
  \label{fig:ordered_GaBiP}
\end{figure}

Figure \ref{fig:ordered_GaBiP} plots the fractional $\Gamma$ character, $G_{\Gamma} (E)$, associated with the GaP VBE states in Ga$_N$Bi$_1$P$_{N-1}$ supercells as the supercell size, $2N$ ($x = N^{-1}$), decreases (increases) from (b) $N = 2048$ ($x = 0.049$\%) to (f) $N = 108$ ($x = 0.926$\%). For reference, we also include in Fig.~\ref{fig:ordered_GaBiP} (a)  the calculated $G_{\Gamma} (E)$ for a pure GaP supercell, in which 100\% of the HH/LH $\Gamma$ character resides on the VBE.

 \vspace{1mm} 

When we replace a single P atom by Bi in a Ga$_{2048}$P$_{2048}$ supercell, we see in Fig.~\ref{fig:ordered_GaBiP} (b) that this gives rise to a four-fold degenerate defect level which lies above, and interacts with, the GaP VBE. The amplitude of the unperturbed GaP VBE states is distributed over two levels, corresponding to $E_{-}$ and $E_{+}$, as expected from the band anti-crossing model of Eq.~(3). We calculate using Eqs. (4) and (5) that the isolated Bi impurity level lies 122 meV above the GaP VBE, and that each impurity state is highly localized, with 31\% of its probability density localized on the Bi atom and the four nearest-neighbor Ga sites. We also calculate that the band anti-crossing interaction $V_{Bi}$ between the Bi-related defect levels, $E_{Bi}$, and the VBE is of magnitude $V_{Bi} = \beta \sqrt{x}$, with $\beta = 1.41$ eV.
 
The calculated magnitude of the interaction parameter $\beta$ in GaBi$_{x}$P$_{1-x}$ is then comparable to that which we have calculated previously \cite{Reilly_1} for GaN$_{x}$P$_{1-x}$ ($\beta = 1.74$ eV) and for GaN$_{x}$As$_{1-x}$ ($\beta = 2.00$ eV).

Having identified in this section that an anti-crossing interaction does indeed occur between Bi-related defect levels and the GaP VBE in GaBi$_{x}$P$_{1-x}$, we now use this understanding to inform our analysis of the valence band structure of GaBi$_{x}$As$_{1-x}$.

 
\subsubsection{Ga$_N$Bi$_1$As$_{N-1}$ supercells}

\begin{figure}
    \includegraphics[scale=0.63]{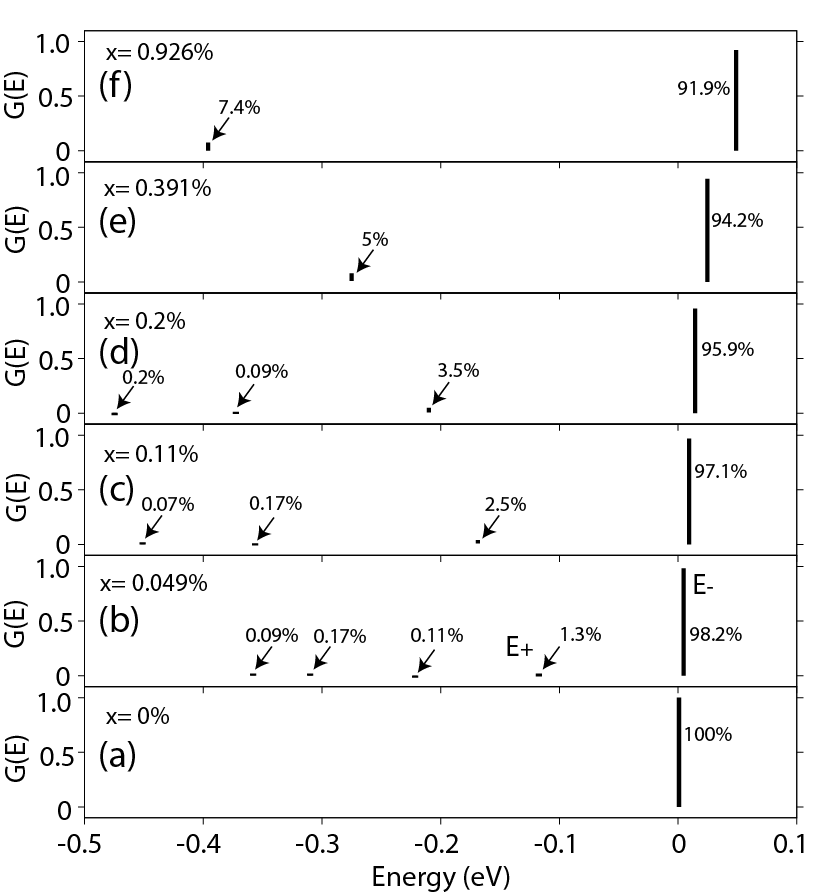}
 \caption{Combined fractional HH/LH $\Gamma$ character for a series of ordered Ga$_{N}$Bi$_{1}$As$_{N-1}$ supercells. The zero of energy is taken at the GaAs VBE.}
  \label{fig:ordered_GaBiAs}
\end{figure}

Figure \ref{fig:ordered_GaBiAs} plots the fractional $\Gamma$ character, $G_{\Gamma} (E)$, associated with the GaAs VBE states in Ga$_N$Bi$_1$As$_{N-1}$ supercells as the supercell size, $2N$ (Bi composition, $x$), decreases (increases) from (b) $N = 2048$ ($x = 0.049$\%) to (f) $N = 108$ ($x = 0.926$\%). For reference, we again include as Fig.~\ref{fig:ordered_GaBiAs} (a)  the calculated $G_{\Gamma} (E)$ for a pure GaAs supercell, in which 100\% of the HH/LH $\Gamma$ character resides on the VBE.

 \vspace{1mm} 
 
 \begin{figure}
 \includegraphics[scale=0.55]{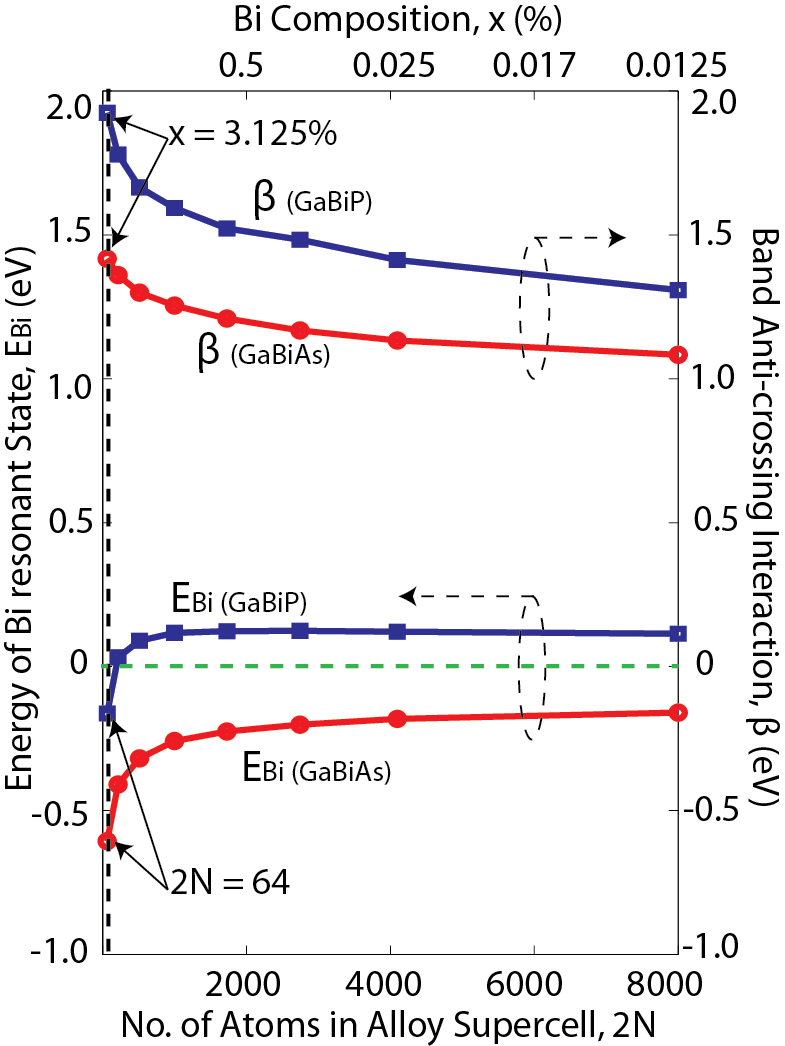}
 \caption{Plot of Bi resonant state energy E$_{Bi}$ and band anti-crossing parameter $\beta$ as a function of the size of GaBi$_{x}$Y$_{1-x}$ (Y=P, As) supercell. Each supercell contains one Bi atom. The horizontal dotted line shows valence band edge energy in GaAs for GaBi$_{x}$As$_{1-x}$ and in GaP for GaBi$_{x}$P$_{1-x}$.}
 \label{fig:resonant_state}
\end{figure}
\vspace{1mm}

We first consider a single substitutional Bi atom in a 4096 atom supercell ($x = 0.049$\%, panel (b) of Fig.~\ref{fig:ordered_GaBiAs}). We see for this case that the HH and LH $\Gamma$ character remains predominantly on the highest alloy valence states, with $98.2$\% of the $\Gamma$ character on these states. Of the remaining $1.8$\%, $1.3$\% is found on a single energy state $\approx$ 117~meV below the GaAs VBE, with the other $0.5$\% spread over a series of lower lying levels. Using Eq.~(4), we identify that the state which mixes with each of the GaAs valence band edge states is a localized Bi resonant defect level, with very similar character to the localized defect state found in GaBi$_{x}$P$_{1-x}$. The Bi state is calculated to be highly localized, with 36\% of its probability density localized on the Bi atom and the four nearest-neighbor Ga sites.

We calculate by evaluating $V_{Bi}$ = $\langle \psi_{Bi} \vert \widehat{H} \vert \psi_{v,0} \rangle$ for an isolated Bi impurity in a 4096 atom supercell that the band anti-crossing interaction $V_{Bi}$ between the Bi-related resonant defect level, $E_{Bi}$, and the GaAs VBE, $E_{v,0}$, is of magnitude $V_{Bi} = \beta \sqrt{x}$, with $\beta = 1.13$ eV. The calculated magnitude of the interaction parameter $\beta$ in GaBi$_{x}$As$_{1-x}$ is then smaller than the value in GaBi$_{x}$P$_{1-x}$, reflecting that there is less difference in electronegativity and size between a Bi and an As atom than there is between a Bi and a P atom.

\begin{figure*}
\includegraphics[scale=0.55]{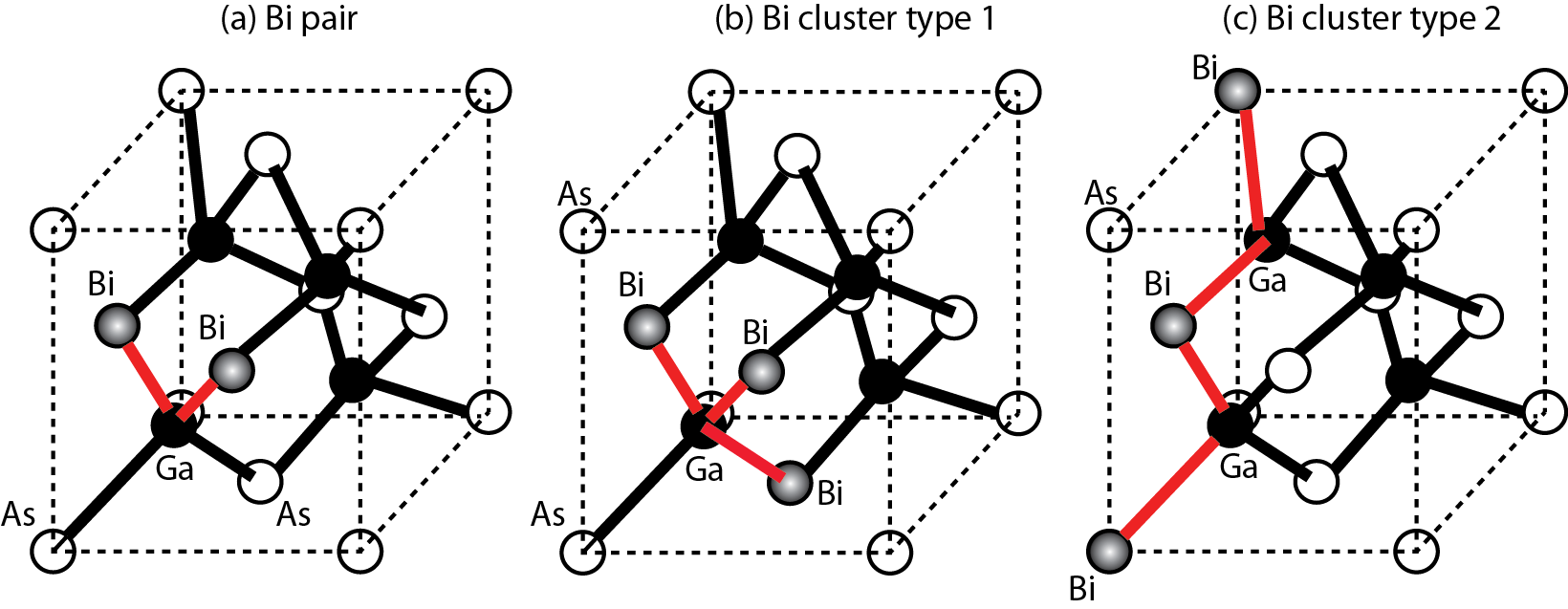}
\caption{Three different Bi nearest-neighbor environments are used in our study of pairing and clustering effects: (a) a single Bi \textquoteleft pair\textquoteright \; (i.e. an isolated Ga atom having two Bi nearest neighbors), (b) a cluster of Bi atoms is placed inside the supercell such that a single Ga atom has three Bi nearest neighbors (\textquoteleft cluster type 1\textquoteright), and (c) a cluster of Bi atoms containing three Bi atoms in the form of a Bi-Ga-Bi-Ga-Bi chain. In this configuration, two Ga atoms have two Bi nearest neighbors (\textquoteleft cluster type 2\textquoteright). Note that the nearest-neighbor environments are depicted here in their unrelaxed configurations. The bonds in the red color indicate the Bi pair and cluster connecting bonds.}
\label{fig:pair_cluster}
\end{figure*}
\vspace{1mm}

When the size of the supercell is decreased from 4096 atoms to 216 atoms (panels (b) to (f) in Fig.~\ref{fig:ordered_GaBiAs}), the interaction of the Bi-related resonant defect state with the GaAs VBE increases, thus pushing the valence band edge, E$_-$, upward in energy. The lower BAC level is pushed downward in energy and its fractional $\Gamma$ character, $G_{\Gamma} (E)$, also increases as a result of the increasing strength of the interaction. Overall, the $\Gamma$ character associated with the lower BAC level is in all cases spread over several supercell levels (clearly visible in panels (b), (c) and (d), but running out of scale to lower energies in panels (e) and (f)), reflecting that the Bi states are resonant with a high density of host matrix valence band states.

Figure~\ref{fig:resonant_state} shows the calculated values of $\beta$ and $E_{Bi}$ for a series of ordered $2N$ atom Ga$_{N}$Bi$_{1}$Y$_{N-1}$ (Y = P, As) supercells, where $N$ is increased from 32 to 4000 atoms, retaining a cubic supercell at each $N$. Both the GaBi$_{x}$As$_{1-x}$ and GaBi$_{x}$P$_{1-x}$ supercells exhibit similar trends. As the supercell size increases, the calculated strength of the band anti-crossing interaction between the Bi impurity state and the GaAs VBE reduces. For larger supercells ($N > 2000$ atoms) we approach the dilute doping limit and the trend stabilises. We also find that the calculated energy of the Bi resonant states, $E_{Bi}$, increases with increasing supercell size. The Bi impurity state is calculated to lie 607~meV below the GaAs VBE in a Ga$_{32}$Bi$_{1}$As$_{31}$ supercell, and its value increases to $-0.18$ eV in a Ga$_{4000}$Bi$_{1}$As$_{3999}$ supercell. This change arises due to the evolution of the GaY (Y = P, As) supercell zone center band structure with increasing supercell size. The GaY supercell zone center states form a complete set of basis states from which to construct the localized Bi states\cite{Lindsay_6}. For the smallest (64 atom) supercell, the highest valence levels folded back to the zone center lie 1.45 eV (1.30 eV) below the VBE in GaAs (GaP). As the supercell size increases, more states get folded back with energies closer to the VBE. These states can be used to construct the Bi localized state energy, thus accounting for the calculated upward shift in the energy $E_{Bi}$ with increasing supercell size.  Based on the trends shown in Fig.~\ref{fig:resonant_state}, we predict values of approximately $-0.18$ eV and $1.13$ eV for $E_{Bi}$ and $\beta$ for GaBi$_x$As$_{1-x}$ alloys in the dilute doping limit, respectively.   


\subsection{Bi pair and cluster states}

Recent experimental investigations using x-ray absorption spectroscopy \cite{Ciatto_1} indicate that Bi atoms in dilute GaBi$_{x}$As$_{1-x}$ alloys tend to form localized pairs and clusters, even at relatively low Bi compositions. The presence of such pairs and clusters of N atoms has been observed in earlier studies on GaN$_{x}$As$_{1-x}$ alloys with dilute N concentrations\cite{Liu_1} and it has been shown that such pairs and clusters significantly affect the electronic structure of the alloy \cite{Kent_1, Masia_1}. We investigate here the impact of Bi pairs and clusters on the electronic structure of GaBi$_{x}$As$_{1-x}$.

We compare the electronic structure of seven different nearest-neighbor environments in a 4096 atom supercell: (i) a Ga$_{2048}$As$_{2048}$ supercell (Bi-free), (ii) Ga$_{2048}$Bi$_{1}$As$_{2047}$ containing an isolated Bi atom, (iii) a Ga$_{2048}$Bi$_{2}$As$_{2046}$ supercell containing two isolated Bi atoms (where each Bi atom has no Bi belonging to either its first, second or third shell of neighboring atoms), (iv) a Ga$_{2048}$Bi$_{2}$As$_{2046}$ supercell containing a pair of Bi atoms (i.e. a Ga atom having two Bi nearest neighbors), (v) a Ga$_{2048}$Bi$_{3}$As$_{2045}$ supercell containing three isolated Bi atoms (with each Bi atom satisfying the same \textquoteleft isolation\textquoteright \; criterion as in (ii)), (vi) a Ga$_{2048}$Bi$_{3}$As$_{2045}$ supercell containing three Bi atoms forming a cluster, such that a single Ga atom has three Bi nearest neighbors (we shall henceforth refer to this nearest-neighbor environment as a cluster of type 1), and (vii) a Ga$_{2048}$Bi$_{3}$As$_{2045}$ supercell containing three Bi atoms forming a Bi-Ga-Bi-Ga-Bi chain, with the two Ga atoms in the chain each having two Bi nearest neighbors (this will be referred to as a cluster of type 2). The nearest-neighbor environments (iv), (vi) and (vii) are depicted in Figs.~\ref{fig:pair_cluster} (a), (b) and (c), respectively.

From the relaxation of the supercell configurations described in (ii) to (vii), we calculate the average relaxed Ga-Bi bond length in each case. For a single Bi atom, the relaxed Ga-Bi bond length is 2.647 \AA. When we introduce a Bi pair, the Bi-Ga bond length to the shared Ga atom reduces to 2.640 \AA. A further reduction in the bond length occurs for Bi clusters. For example, the average Bi-Ga bond length to the shared Ga atoms is 2.640 \AA for a cluster of type 2. These trends are similar to the ones reported by experimental study\cite{Ciatto_1}, where a reduction in the bond lengths is observed for pairs and clusters. The x-ray absorption measurements indicated average bond length d$_{av}$ values of 2.627 \AA, 2.621 \AA, and 2.619 \AA for isolated, pairs, and clusters of Bi atoms.    

\begin{figure}
  \includegraphics[scale=0.63]{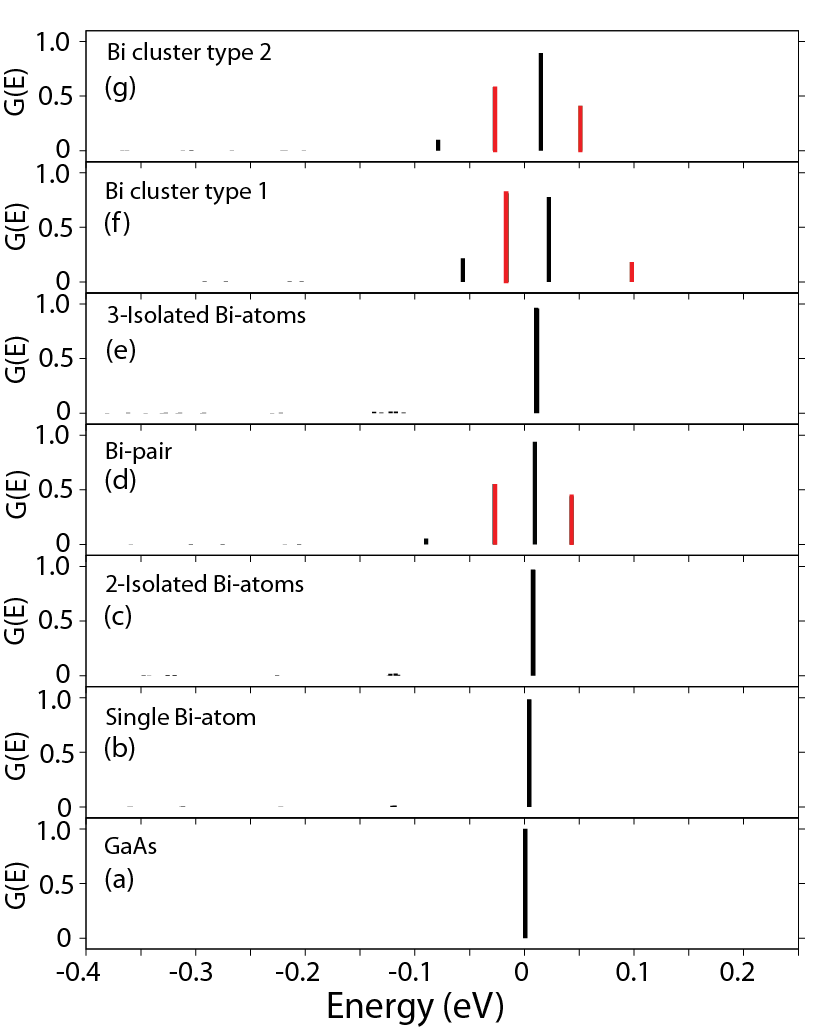}
  \caption{Combined fractional HH/LH $\Gamma$ character for a series of disordered Ga$_{2048}$Bi$_{M}$As$_{2048-M}$ supercells, for several different nearest-neighbor environments. The zero of energy is taken at the GaAs VBE.}
  \label{fig:FGC_pair_cluster}
\end{figure}

Figure~\ref{fig:FGC_pair_cluster} shows the calculated fractional GaAs VBE $\Gamma$ character for all seven of these nearest-neighbor environments. We recall in panel (b) that inclusion of a single Bi atom results in the mixing of a small amount ($\approx 2$\%) of the GaAs VBE $\Gamma$ character into alloy states in the VB. Similar weak interactions can be seen in panels (c) and (e), for the cases where two and three isolated Bi atoms are incorporated in the alloy. In these cases, a large fraction (96\% in panel (c) and 95\% in panel (e)) of $G_{\Gamma} (E)$ again resides on the highest alloy valence state, with the remaining 4-5\% distributed over a number of lower valence levels. 

Turning to the Bi pair in panel (d), the $\Gamma$ character is in this case distributed over four states, which form two interacting, anti-crossing pairs (indicated by red and black bars). One of the pairs shown with the black color bars ($\approx$ 5.3\% weight at -90~meV and $\approx$ 94\% at 10~meV) behaves in a very similar way to the pair of states associated with isolated Bi atoms in Fig.~\ref{fig:FGC_pair_cluster} (b) and (c), while the second pair (red bars) behaves quite differently, with $\approx$ 55\% $\Gamma$ character on the lower state at -27~meV and $\approx$ 44\% $\Gamma$ character on the upper state at 43~meV, associated with a Bi pair-related defect level  just above the host matrix VBE. We can understand the behavior of these defect states based on interactions between the resonant states associated with two isolated Bi atoms. The formation of the Bi pair breaks the $T_d$ symmetry about each Bi site, lifting the four-fold degeneracy of the Bi-related localized states to give a pair of two-fold degenerate states, one of which has a small interaction and the other a large interaction with Bi states on the neighboring Bi site. The interaction between the two Bi states with large overlap then gives rise to the defect level which we calculate to be in the gap, similar to the behavior of the N-N pair state in GaAs\cite{Masia_1, Liu_1}, while the interaction between the other two sets of Bi states gives the resonant level calculated to be 90~meV below the host matrix VBE. 

\vspace{1mm}

A similar analysis holds for the two types of Bi cluster considered. The type 2 cluster (chain of 3 Bi atoms) has a very similar spectrum to the Bi pair, with a stronger interaction between the lower defect and the VBE states than in the pair case, and with the higher defect state being slightly deeper in the energy gap. We estimate that the energy level of the Bi defect state ($E_{Bi}$) is 10 meV, 8 meV and 17 meV above the GaAs VBE for the Bi pair and the clusters of type 1 and 2, respectively. The lower defect state is shifted to -48 meV in the presence of a type 1 cluster, where one Ga atom has 3 Bi neighbors. Overall, the results follow similar trends to those found previously for N-related defect clusters in GaAs, where the defect levels tended to move deeper into the GaAs energy gap with increasing cluster size.

Table \ref{tab:table5} summarises the calculated effect of the different Bi configurations on the lowest conduction band energy, $E_{c}$, the top two valence bands, $E_{v1}$ and $E_{v2}$, and the spin-split-off band, $E_{SO}$. Also shown in parentheses below each energy are the calculated change in energy for the band (per \% Bi incorporation, relative to the corresponding GaAs band edge) as well as the overlap of the state with the corresponding unperturbed GaAs state. We  see from these calculations that $E_{c}$ and $E_{SO}$ are both largely insensitive to the nearest-neighbor environment in the Bi-containing supercells. The conduction band exhibits a decrease of 28~meV per \% Bi, which is linear in Bi composition, and retains a 99.9\% overlap with the GaAs CBE for all Bi configurations considered. Thus, the variation in the conduction band and spin-split-off energies with increasing Bi composition can be well explained in terms of conventional alloy models, irrespective of disorder in the form of Bi pairs and clusters in the alloy. We note that unexpectedly large values have been deduced for the conduction band edge mass in GaBi$_x$As$_{1-x}$ alloys for $x \leq 6\%$\cite{Pettinari_2}. Further studies are required to identify the cause of the anomalous CBE mass values deduced experimentally: our calculations indicate that it should be possible to describe the CBE mass in randomly disordered GaBi$_{x}$As$_{1-x}$ using conventional alloy models.


\subsection{Electronic structure of disordered Ga$_{2048}$Bi$_{M}$P$_{2048-M}$ supercells}

Having established the character of isolated Bi single, pair and cluster impurity states in sections III-A and III-B, we now investigate the electronic structure of randomly disordered GaBi$_{x}$P$_{1-x}$ alloys, as we increase the number ($M$) of Bi atoms in a Ga$_{2048}$Bi$_{M}$P$_{2048-M}$ supercell. In order to trace the evolution of the highest valence states, we diagonalise the $sp^{3}s^{*}$ Hamiltonian for the supercell and use the resulting alloy states to calculate the fractional $\Gamma$ character, $G_{\Gamma} (E)$, for the VBE (Eqs. (7) and (8)). 

\begin{figure}
    \includegraphics[scale=0.63]{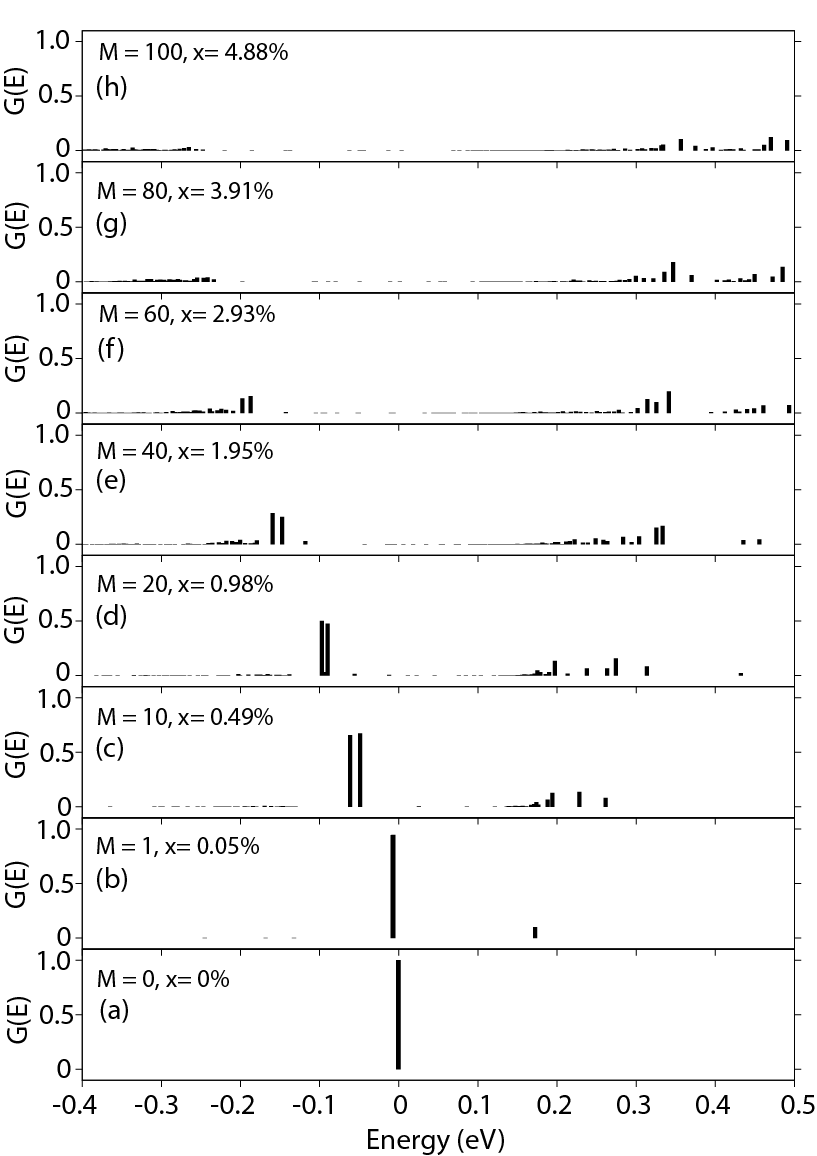}
 \caption{Combined fractional HH/LH $\Gamma$ character for a series of disordered Ga$_{2048}$Bi$_{M}$P$_{2048-M}$ supercells, for several Bi compositions, $x = \frac{M}{2048}$. The zero of energy is taken at the GaP VBE.}
  \label{fig:FGC_GaBiP}
\end{figure}

The calculated $G_{\Gamma} (E)$ spectra for a range of Bi compositions are shown in Fig.~\ref{fig:FGC_GaBiP}. We see in panel (b) for a single Bi atom in the supercell ($M = 1$, $x = 0.05$\%) that the amplitude of the unperturbed GaP VBE states is distributed over two levels, corresponding to $E_{-}$ and $E_{+}$, as previously discussed in section III-A.

 \vspace{1mm}
 
 \begin{table*}
\caption{\label{tab:table5}Calculated effect of Bi isolated, pair, and clusters of atoms on the lowest conduction band, the highest two valence bands, and the spin-split-off band edges of Ga$_{2048}$Bi$_M$As$_{2048-M}$ supercells, including energy of given state and in parentheses  the calculated change in energy for the band (per \% Bi incorporation, relative to the corresponding GaAs band edge) as well as the overlap of the state with the corresponding unperturbed GaAs state.}
\begin{ruledtabular}
\begin{tabular}{ccccccc}
\textbf{Alloy Composition} & \textbf{$x$(\%)} & \textbf{E$_{c1}$} (eV) & \textbf{E$_{v1}$} (eV) &\textbf{ E$_{v2}$} (eV) & \textbf{E$_{so}$} (eV)  \\
 & & (meV/\%Bi, \%) & (meV/\%Bi, \%) & (meV/\%Bi, \%) & (meV/\%Bi, \%) \\ 
\hline \\
\textbf{Unperturbed GaAs} & \textbf{0} & \textbf{1.519} & \textbf{0.000773} & \textbf{0.000773} & \textbf{-0.3518} \\ 
& & (0, 100) & (0, 100) & (0, 100) & (0, 100) \\ \hline \\ 
\textbf{Single Bi-atom} & \textbf{0.049} & \textbf{1.5176} & \textbf{0.00456} & \textbf{0.004359} & \textbf{-0.3521} \\ 
& & (28.57, 99.97) & (77.28, 98.44) & (73.18, 98.29) & (7.06, 99.95) \\ \hline \\
\textbf{2-isolated Bi-atoms} & \textbf{0.098} & \textbf{1.5163} & \textbf{0.008106} & \textbf{0.007746} & \textbf{-0.35244} \\ 
& & (27.55, 99.93) & (74.82, 96.92) & (71.15, 97.15) & (6.53, 99.94) \\ \\
\textbf{Bi-pair} & \textbf{0.098} & \textbf{1.5162} & \textbf{0.04298} & \textbf{0.009488} & \textbf{-0.35244} \\ 
& & (28.57, 99.93) & (430.68, 44) & (88.92, 94.02) & (6.62, 99.34) \\ \hline \\
\textbf{3-isolated Bi-atoms} & \textbf{0.146} & \textbf{1.5149} & \textbf{0.01146} & \textbf{0.01078} & \textbf{-0.35273} \\ 
& & (28.08, 99.9) & (73.2, 95.8) & (68.51, 96.44) & (6.33, 99.75) \\ \\
\textbf{Bi-cluster Type 1} & \textbf{0.146} & \textbf{1.5148} & \textbf{0.09803} & \textbf{0.0223} & \textbf{-0.3526} \\ 
& & (28.76, 99.88) & (666.14, 18.14) & (147.44, 77.73) & (5.6, 99.75) \\ \\
\textbf{Bi-cluster Type 2} & \textbf{0.146} & \textbf{1.5149} & \textbf{0.05117} & \textbf{0.01493} & \textbf{-0.35273} \\
& & (28.08, 99.9) & (345.18, 41.06) & (96.96, 89.34) & (6.35, 99.64) \\ \\
\end{tabular}
\end{ruledtabular}
\end{table*}

In the supercells used in calculations containing $M \geq 10$ Bi atoms ($x \geq 0.49$\%) we see that the calculated splitting between the $E_{-}$ and $E_{+}$ levels does not increase smoothly with increasing Bi composition. In addition, because of the reduced symmetry in these supercells and the inclusion of Bi pairs and clusters, the two highest valence states are no longer degenerate and we see an overall broadening of the spectra.

The pair and cluster states introduce higher energy defect states, lying above the isolated Bi states in the energy gap, as can be seen at higher Bi composition in Fig.~\ref{fig:FGC_GaBiP}. The Bi defect state distribution in GaBi$_{x}$P$_{1-x}$ then has a similar form to that observed in GaN$_{x}$P$_{1-x}$, where isolated N defect states are just below the GaP CBE, with defects states associated with N pairs and other cluster levels lying deeper in the energy gap \cite{Harris_1}.

We can see from Fig.~\ref{fig:FGC_GaBiP} that there is  a general anti-crossing interaction between the valence band states and the Bi defect states in the gap, with an increasing separation between the VB-related states (below 0 eV) and the Bi states in the gap ($\gtrsim 0.2$~eV) with increasing Bi composition. However, because the gap defect levels are distributed over a range of energies, it is not possible to describe the band structure close to the energy gap using the two-level band anti-crossing model of Eqs. (1) and (3).  We showed previously for GaN$_{x}$P$_{1-x}$ that it is possible to treat the anti-crossing interaction between the GaP host matrix $\Gamma$ CBE and a distribution of N-related defect levels  by treating explicitly the interaction between the GaP $\Gamma$ conduction state and the full distribution of N defect levels\cite{Harris_1}. We see from Fig.~\ref{fig:FGC_GaBiP} that this should also be the case in GaBi$_{x}$P$_{1-x}$. When the supercell only included isolated Bi atoms then a clear anti-crossing interaction was observed in Fig.~\ref{fig:ordered_GaBiP}, in agreement with Eq. (3). However, when the calculation includes a wider distribution of Bi environments (which is expected experimentally even for $x < 1$\%) then the 2-level BAC model is no longer able to describe the observed evolution of the VBE states. We expect that  a model similar to that used for GaN$_{x}$P$_{1-x}$ could also be used to describe the evolution of the band structure of GaBi$_{x}$P$_{1-x}$. We do not however pursue that further here.


\subsection{Electronic structure of disordered Ga$_{2048}$Bi$_{M}$As$_{2048-M}$ supercells}


\subsubsection{Evolution of GaBi$_{x}$As$_{1-x}$ valence states with $x$}

\begin{figure}
 \includegraphics[scale=0.63]{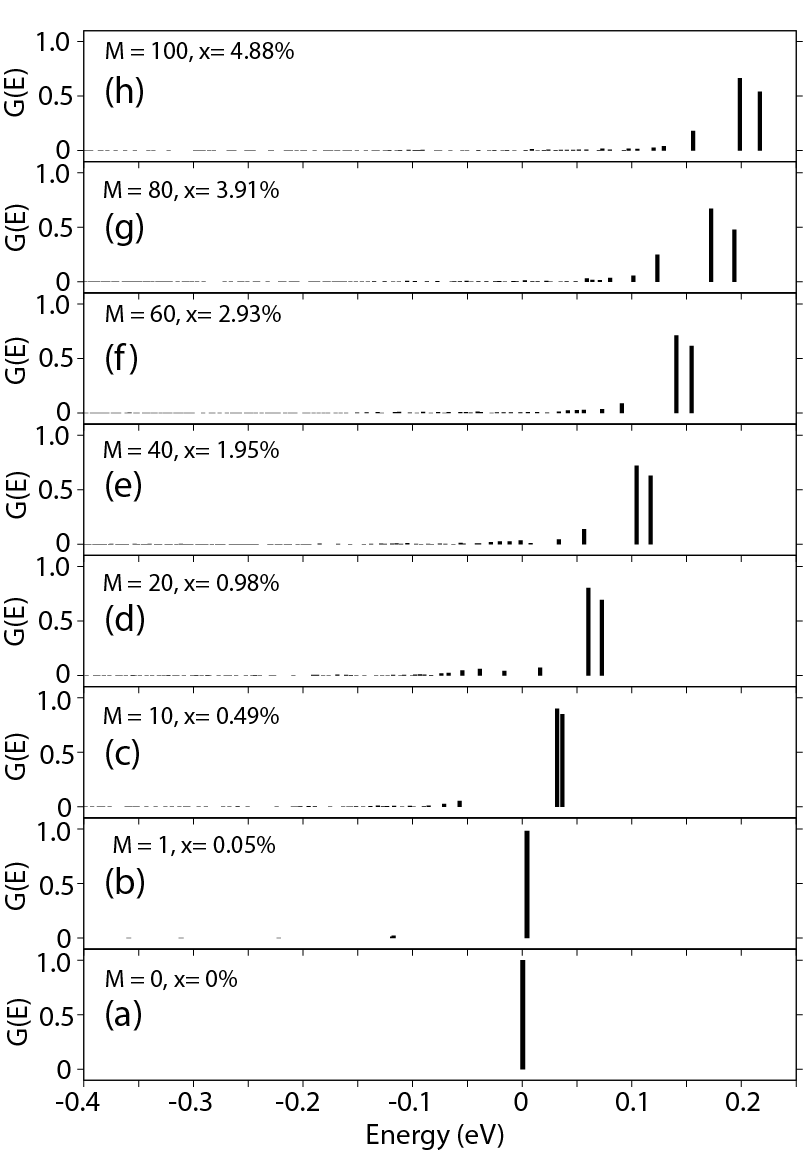}
 \caption{Combined fractional HH/LH $\Gamma$ character for a series of disordered Ga$_{2048}$Bi$_{M}$As$_{2048-M}$ supercells for several Bi compositions, $x = \frac{M}{2048}$. The zero of energy is taken at the GaAs VBE.}
 \label{fig:FGC_GaBiAs}
\end{figure}

Figure~\ref{fig:FGC_GaBiAs} shows the calculated evolution of $G_\Gamma(E)$ for the highest valence states in a series of Ga$_{2048}$Bi$_{M}$As$_{2048-M}$ supercells, where the supercell energy spectrum is in each case projected onto the unperturbed GaAs valence band edge heavy-hole (HH) and light-hole (LH) wave functions (see Eqs. (7) and (8)). To enable comparison with the results of our GaBi$_{x}$P$_{1-x}$ calculations, we consider identical Bi distributions within the supercells, with all P atoms replaced by As. For reference, the lowest panel, (a), shows the calculated $G_{\Gamma} (E)$ for an unperturbed Ga$_{2048}$As$_{2048}$ supercell, which is simply a single peak of unit amplitude, centered at the location of the Ga$_{2048}$As$_{2048}$ HH/LH energy levels.

\begin{figure}
 \includegraphics[scale=0.7]{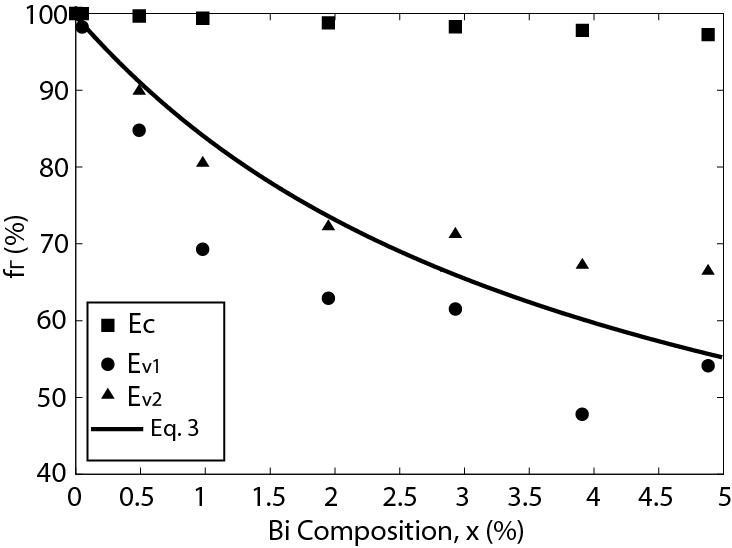}
 \caption{ Plot of fractional $\Gamma$ character, $f_{\Gamma}$ (see Eq. 8), of the lowest conduction band (E$_{c}$) and the highest two valence band (E$_{v1}$ and E$_{v2}$) edges as a function of Bi composition. The solid line shows the valence band $f_{\Gamma}$ character calculated using parameters in Eq.~3 extracted from tight-binding calculations of supercells containing isolated Bi atoms. }
 \label{fig:Overlaps_Ec_Ev}
\end{figure}

We observe in Fig.~\ref{fig:FGC_GaBiAs} that the calculated VBE energy shifts upwards with increasing Bi composition, $x$, and that the overall $\Gamma$ character of the valence band maximum state decreases with increasing ${x}$, as expected from the BAC model of Eq. (3). The solid circles and triangles in Fig.~\ref{fig:Overlaps_Ec_Ev} show how the calculated fractional $\Gamma$ character, $f_{\Gamma,v} = \vert \langle \psi_{v,0} \vert \psi_{v,1} \rangle \vert^{2}$, of each of the two highest valence states decreases with increasing Bi composition in the supercells considered.

The solid line in Fig.~\ref{fig:Overlaps_Ec_Ev} shows the expected variation in $\Gamma$ using Eq. (3) with: $E_{HH/LH} = 1.03 \; x$ eV, $E_{Bi} = -0.18$ eV and $V_{Bi} = \beta \sqrt{x}$, with $\beta = 1.13$ eV. It can be seen that a good fit is obtained at low $x$, with the quality of the fit decreasing at higher values of $x$, as we move into a disordered regime in the alloy.

\vspace{1mm}

\vspace{1mm}

\begin{figure}
 \includegraphics[scale=0.63]{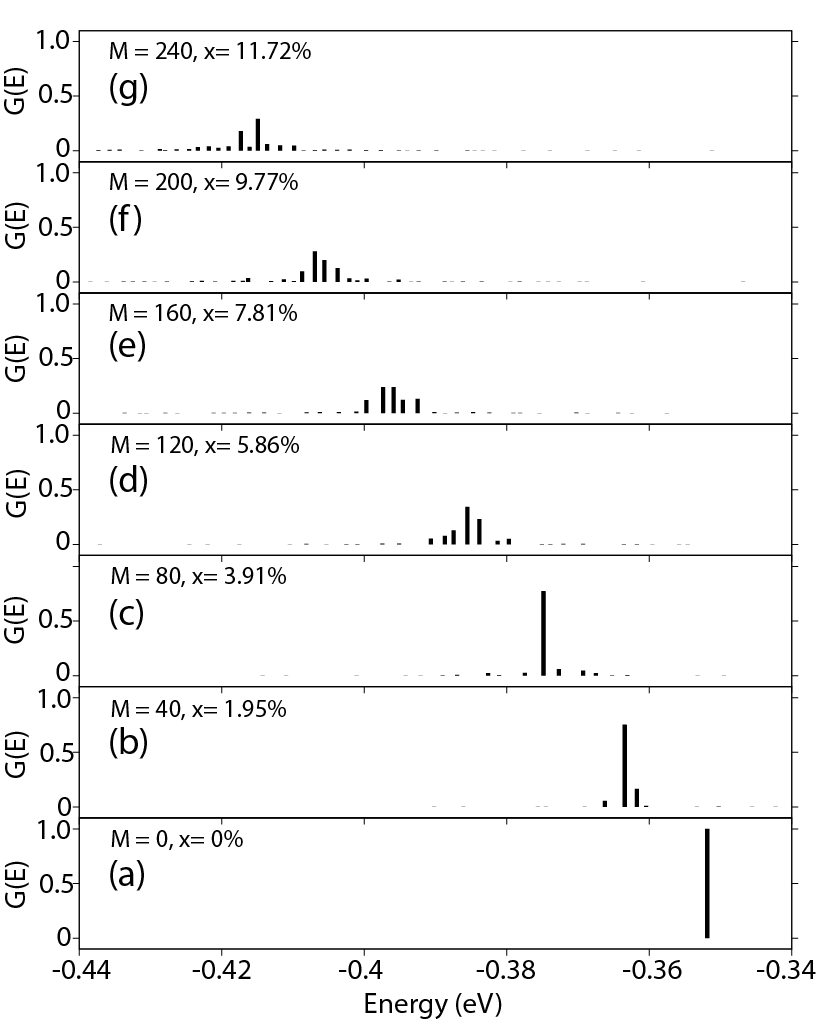}
 \caption{Combined fractional SO $\Gamma$ character for a series of disordered Ga$_{2048}$Bi$_{M}$As$_{2048-M}$ supercells for several Bi compositions, $x = \frac{M}{2048}$.}
 \label{fig:FGC_SO_GaBiAs}
\end{figure}
\vspace{1mm}

A resonant feature associated with the upper eigenvalue, $E_{+}$, of the Hamiltonian (1) has been observed in PR measurements \cite{Perkins_1, Klar_1} on GaN$_{x}$As$_{1-x}$ for $x > 0.2$\%, giving a relatively sharp feature until $x$ reaches approximately 3\%, beyond which composition the feature broadens and weakens, when the resonant state becomes degenerate with the $L$-related CB levels \cite{Lindsay_3}. It has not proved possible to observe such a feature at any composition in GaBi$_{x}$As$_{1-x}$, and this has led to controversy as to whether there is a VB-BAC interaction present in the alloy\cite{Deng_1}. It can be seen from Fig.~\ref{fig:FGC_GaBiAs} that the GaAs VBE character in GaBi$_{x}$As$_{1-x}$ is mixed into a wide range of the alloy valence states so that, unlike in the GaBi$_{x}$P$_{1-x}$ case, no single state can be identified as the $E_{+}$ level in GaBi$_{x}$As$_{1-x}$. This strong mixing is to be expected, because the width of the resonant level is proportional to the density of states\cite{Fano_1, Fahy_1}. The large density of valence band states then explains the failure to observe an $E_{+}$ level in GaBi$_{x}$As$_{1-x}$ PR spectra. 

\subsubsection{Band gap and spin-orbit-splitting energies versus $x$}

Figure~\ref{fig:FGC_SO_GaBiAs} shows how the fractional $\Gamma$ character of the SO band edge evolves with increasing Bi composition. As the Bi composition increases, the $\Gamma$ character spreads out over several states and it becomes unclear to choose a particular energy for $E_{SO}$. The values of the $E_{SO}$ energies from panels (b) to (g) of Fig.~\ref{fig:FGC_SO_GaBiAs} are obtained by taking a weighted average of the energies with their corresponding $\Gamma$ character as well as by picking the energies with the highest $\Gamma$ character. The values from the two methods come out to be very close (within 0.1~meV of each other) and we deduce that both methods are equally good in this case. The Bi-induced disorder leads to a broadening of the projected $\Gamma$ character, $G_{\Gamma} (E)$, with increasing Bi composition, but no anti-crossing behavior is observed in this broadened spectrum. Hence, the shift and broadening of the SO state projection is attributable to conventional alloying  and disordering effects. For this reason we conclude that our calculations do not support the existence of a Bi resonant state interacting with the SO band, as introduced in the $\textbf{k} \cdot \textbf{p}$ Hamiltonian of Alberi \emph{et al.} \cite{Alberi_1}. 

\begin{figure}
 \includegraphics[scale=0.55]{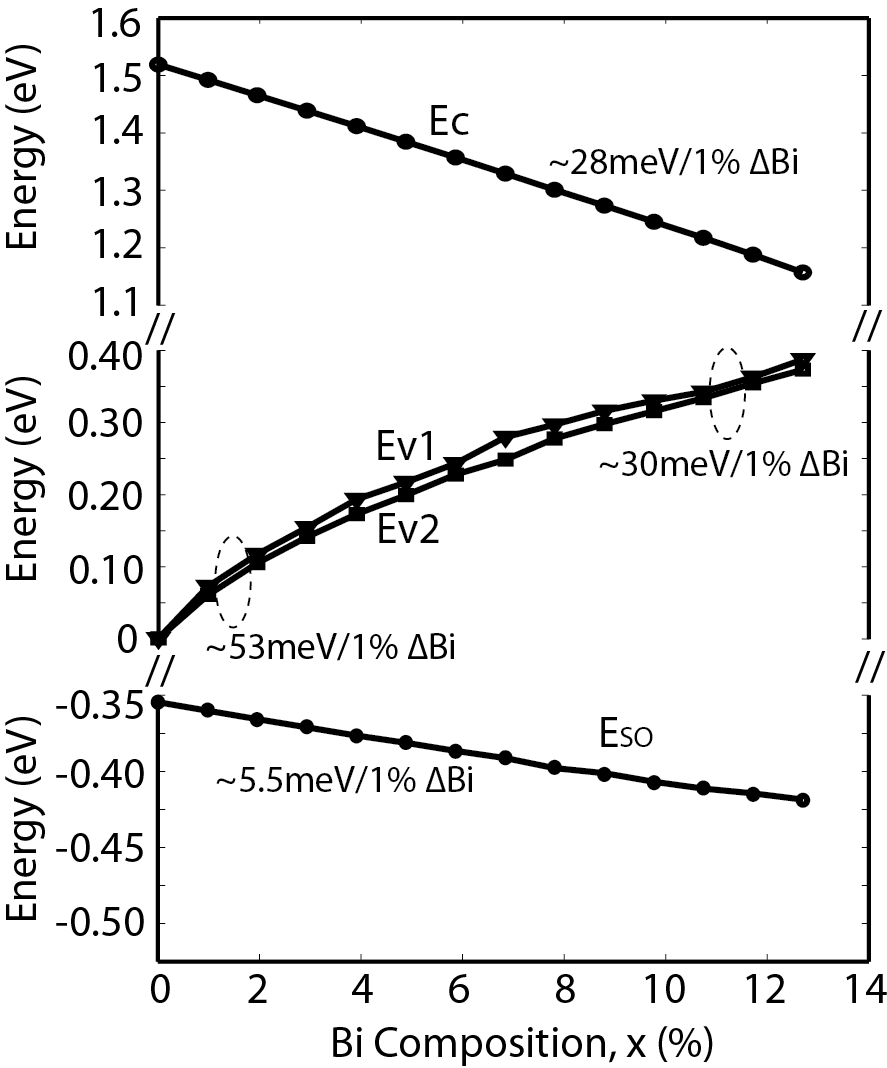}
 \caption{The lowest conduction band edge $E_c$, the highest two valence band edges $E_{v1}$, $E_{v2}$, and the spin-split-off band $E_{so}$ energies are plotted as a function of Bi composition in GaBi$_{x}$As$_{1-x}$ supercells. The conduction band $E_c$ edge decreases with a linear slope, changing by about {28~meV} for every {1}\% increase in the Bi composition. The highest two valence band edges $E_{v1}$ and $E_{v2}$ exhibit a non-linear change. For small Bi composition, the slope is large {53}~meV/{1}\% Bi. For large Bi compositions, the slope decreases to about {30}~meV/{1}\% Bi. The spin-split-off band $E_{so}$ also decreases following a nearly linear slope of {5.5}~meV/{1}\% Bi.}
 \label{fig:band_edges}
\end{figure}

Figure ~\ref{fig:band_edges} shows the calculated variation of the lowest conduction, the two highest valence, and the spin-split-off band edge energies with increasing Bi composition $x$ in the {4096} atom GaBi$_{x}$As$_{1-x}$ supercells. It can be seen that the calculated conduction band edge energy shifts down linearly by about 28 meV per \% of Bi replacing As. The solid squares in Fig.~\ref{fig:Overlaps_Ec_Ev} show that $E_c$ retains $\geq$ 97\% GaAs conduction band $\Gamma$ character for $x \leq {5} \%$. With a linear decrease in the energy of $E_c$ and the $\Gamma$ character remaining close to 100\%, we conclude that the evolution of the conduction band states is also best described using a conventional alloy model.

\vspace{1mm}

An important feature of GaBi$_{x}$As$_{1-x}$ alloys is that, as the Bi composition increases, the decreasing band gap $E_g$ = $E_c$ - $E_{v1}$  and the increasing spin-orbit splitting energy $\Delta_{SO}$ = $E_{v1}-E_{SO}$ results in a crossing of these two energies, with $E_g < \Delta_{SO}$ at higher Bi compositions. It has been suggested that this should lead to a suppression of the dominant CHSH Auger recombination mechanism in GaBi$_{x}$As$_{1-x}$ at higher Bi compositions, which could be highly beneficial for the design of high efficiency lasers and semiconductor optical amplifiers (SOAs) operating at telecommunication and at longer wavelengths\cite{Sweeney_1}. Recent PR measurements \cite{Batool_1} have confirmed a crossing of the energy gap and spin-orbit-splitting energies in bulk GaBi$_x$As$_{1-x}$ alloys for compositions above about $x$=10.5$\%$. 

\begin{figure}
 \includegraphics[scale=0.6]{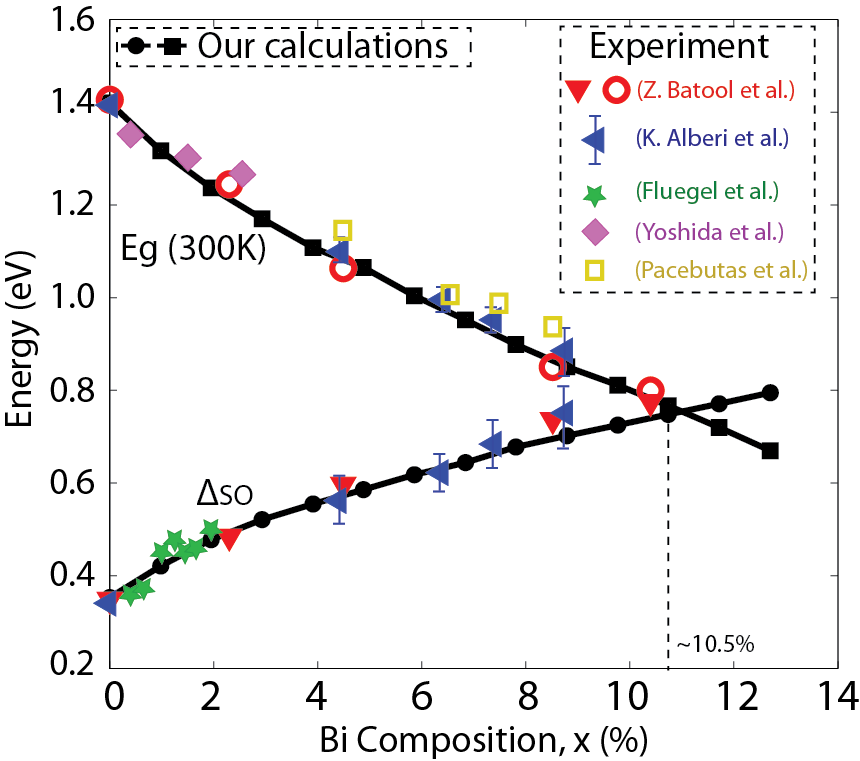}
 \caption{The band gap $E_g$ and spin-orbit-splitting $\Delta_{SO}$ energies plotted as a function of Bi composition in 4096 atom GaBi$_{x}$As$_{1-x}$ alloy supercells. The solid circles and squares show the calculated data from our model. The triangles and open circles indicate the measured PR data from experiment (Z. Batool \textit{et al.}\cite{Batool_1}]). Four other sets of experimental data are also included for broader comparison. }
 \label{fig:eg_so}
\end{figure}

Figure ~\ref{fig:eg_so} plots the calculated band gap $E_g$ and spin-orbit splitting $\Delta_{SO}$ as a function of Bi composition in {4096} atom GaBi$_{x}$As$_{1-x}$ supercells. The calculations were undertaken using the low temperature GaBi$_{x}$As$_{1-x}$ band parameters presented here ($E_g$(GaAs)$ = 1.519$ eV), and the CBE energy has then been shifted down by 92 meV for all Bi compositions, to fit to the GaAs energy gap at room temperature. Each data point shown in the figure is calculated by taking the average of three different random configurations of GaBi$_{x}$As$_{1-x}$. The results presented indicate that the decreasing band gap $E_g$  and the increasing spin-orbit splitting do indeed lead to a crossover of $E_g$ and $\Delta_{SO}$ at about 10.5\% Bi composition, in agreement with PR measurements\cite{Sweeney_1}. For a broader comparison with other experiments, we have also included four sets of experimentally measured band gap and spin-orbit splitting values\cite{Alberi_1, Fluegel_1, Yoshida_1, Pacebutas_1} in Fig.~\ref{fig:eg_so} along with our calculated values. Our results exhibit a close match with the experimental data points over the entire composition range considered, confirming the validity of our model.

\vspace{1mm}


\section{Conclusions}

In summary, we have developed a  model based on the valence force field (VFF) and $sp^{3}s^{*}$ tight binding methods to study the dilute bismide alloys GaBi$_x$As$_{1-x}$ and GaBi$_x$P$_{1-x}$.  Our results  confirm that the evolution of the highest valence states in GaBi$_x$Y$_{1-x}$ (Y = P, As) can be described using a band anti-crossing model. This has, until now, proved difficult to confirm experimentally because one of the key BAC features, the E$_{+}$ state observed in GaN$_{x}$As$_{1-x}$, is absent in GaBi$_{x}$As$_{1-x}$. We showed that this higher-lying E$_+$ level which has been identified in GaN$_{x}$As$_{1-x}$ cannot be observed in GaBi$_{x}$As$_{1-x}$, due to the resonant levels being strongly broadened because of their interaction with the large density of valence states. In addition, we find that the presence of Bi pairs and clusters in a randomly disordered alloy introduce further defect levels, which also contribute to the suppression of the $E_+$ feature. We calculate that the experimentally observed strong band gap reduction is only partially explained through the valence band BAC interaction. We find that the conduction band edge energy also decreases with increasing Bi composition, and that both band edge shifts contribute approximately equally (at large Bi compositions) to the reduction in the band gap energy of unstrained GaBi$_{x}$As$_{1-x}$ alloys. The computed dependence of band gap on Bi composition closely matches with  experimental measurements. We calculate that the band gap and spin-orbit-splitting energies cross each other at ~$10.5$\% Bi composition in GaBi$_{x}$As$_{1-x}$, consistent with recent experimental measurements\cite{Sweeney_1}. It has been suggested that this crossover may eliminate the dominant CHSH Auger recombination loss mechanism in longer wavelength photonic devices. Overall, we conclude that dilute bismide alloys open up new opportunities for band structure engineering and its application, with significant potential benefits for GaAs-based optical devices.


\section{Acknowledgements}

This work was, in part, financed by the European Union project BIANCHO (FP7-257974) and by Science Foundation Ireland (06/IN.1/I90).  C. A. Broderick acknowledges financial support from the Irish Research Council for Science, Engineering and Technology under the Embark Initiative (RS/2010/2766).


\end{document}